\documentclass[12pt]{article}

\pdfoutput=1
\usepackage{jheppub}
\usepackage{graphicx,epsfig,amsmath,amssymb,mathtools}
\usepackage{subcaption}
\usepackage{comment}
\usepackage{cleveref}
\usepackage{multicol}
\usepackage{multirow}
\graphicspath{{feyndiags/}}
\usepackage[amssymb]{SIunits}
\Crefname{figure}{Fig.}{Figs.} 

\newcommand{\muR}{\ensuremath{\mu_{\rm{R}}}}
\newcommand{\muF}{\ensuremath{\mu_{\rm{F}}}}

\def\be{\begin{equation}}
\def\ee{\end{equation}}
\def\bea{\begin{eqnarray}}
\def\eea{\end{eqnarray}}

\newcommand{\powhegbox}{{\tt{POWHEG-BOX-V2}}}

\newcommand{\gosam}{\textsc{GoSam}{}}
\newcommand{\form}{{\tt FORM}}

\newcommand{\ninja}{{\tt Ninja}}
\newcommand{\avholo}{{\tt OneLOop}}
\newcommand{\qcdloop}{{\tt QCDloop}}

\newcommand{\cg}{c_{g}}
\newcommand{\kg}{\kappa_{g}}

\newcommand{\ct}{c_{t}}

\newcommand{\pth}{p_{T,H}}
\newcommand{\pthmin}{p_{T,H}^{\mathrm{min}}}

\title{Probing anomalous Higgs boson couplings in Higgs plus jet production at NLO QCD with full $m_t$-dependence}

\author[a]{Benjamin~Campillo~Aveleira,}
\author[a]{Gudrun~Heinrich,}
\author[a]{Matthias~Kerner,}
\author[a]{Lucas~Kunz}

\affiliation[a]{Institute for Theoretical Physics, Karlsruhe Institute
of Technology, Wolfgang-Gaede-Str.~1, 76131 Karlsruhe, Germany}

\emailAdd{benjamin.campillo@kit.edu}
\emailAdd{gudrun.heinrich@kit.edu}
\emailAdd{matthias.kerner@kit.edu}
\emailAdd{lucas.kunz@partner.kit.edu}

\preprint{{\small  KA-TP-17-2024, P3H-24-058}}

\abstract{
 We present  NLO QCD results for Higgs boson production in association with  one jet, including  anomalous Higgs-top  and  effective
  Higgs-gluon couplings within non-linear Effective Field Theory (HEFT), as well as the full top quark mass dependence.
  We provide differential results for the Higgs boson transverse
  momentum spectrum at $\sqrt{s}=13.6~\tera \electronvolt$ and discuss
  the effects of the anomalous couplings.
}


\begin{document}

                       \maketitle

\section{Introduction}

Run 2 of the LHC was very successful in establishing the Higgs boson
couplings to vector bosons and heavy fermions. With Run 3 and the
high-luminosity upgrade the precision of the coupling measurements
will further increase. Therefore it is important to have good control
over the theoretical uncertainties, in particular to distinguish beyond the Standard Model (BSM)
effects from effects due to insufficient higher order corrections.

It has been noticed some time
ago~\cite{Harlander:2013oja,Azatov:2013xha,Banfi:2013yoa,Grojean:2013nya,Schlaffer:2014osa,Buschmann:2014twa,Dawson:2014ora,Buschmann:2014sia,Langenegger:2015lra,Azatov:2016xik,Lindert:2018iug,DiMicco:2019ngk}
that the Higgs boson $p_T$-spectrum is an important observable to constrain both the Yukawa-couplings to top
(and bottom) quarks as well as effective gluon-Higgs couplings due to
interactions with unknown particles at higher scales, leading to operators
that are not present in the Standard Model (SM). The impact of such
effective operators on the $\pth$ spectrum in Higgs+jet production,
including next-to-leading Order (NLO) QCD corrections in the heavy top limit (HTL), has
been investigated in Refs.~\cite{Maltoni:2016yxb,Grazzini:2016paz,Grazzini:2018eyk,Battaglia:2021nys,Maltoni:2024dpn,DiNoi:2024ajj},
where the analysis of Refs.~\cite{ Battaglia:2021nys,Maltoni:2024dpn,DiNoi:2024ajj} also includes investigations of effects due to the renormalisation group running of the Wilson coefficients.
 
\vspace*{3mm}

In the SM, the leading order (LO) one-loop amplitude has been calculated in Ref.~\cite{Baur:1989cm}, NLO results beyond the HTL were first obtained for the high
transverse momentum region~\cite{Melnikov:2016qoc,Lindert:2018iug}. The NLO result  for general kinematics has first been obtained numerically~\cite{Jones:2018hbb,Becker:2020rjp,Neumann:2018bsx}, the top quark mass effects have been studied in detail in Ref.~\cite{Chen:2021azt}.
Numerical results for the two-loop amplitude for Higgs plus jet production with full top quark mass dependence also have been calculated in Ref.~\cite{Czakon:2021yub} in the framework of the full next-to-next-to-leading order (NNLO) corrections to inclusive Higgs production, a study of the Higgs boson transverse momentum spectrum through the combination of this calculation with a parton shower has been presented in Ref.~\cite{Niggetiedt:2024nmp}.\\
On the analytic side, compact analytic expressions for the one-loop amplitudes for Higgs~+~4 parton scattering, providing an efficient way to calculate the NLO real radiation contributions, have been presented in Ref.~\cite{Budge:2020oyl}.
After the relevant two-loop master integrals have become available~\cite{Bonciani:2016qxi,Bonciani:2019jyb,Frellesvig:2019byn},  the full NLO calculation based on analytic results for the two-loop integrals has been completed in Ref.~\cite{Bonciani:2022jmb}, including both top and bottom masses as well as a comparison of the on-shell and $\overline{\mathrm{MS}}$ schemes to renormalise the top quark mass. Two-loop bottom quark mass effects also have been studied in Ref.~\cite{Pietrulewicz:2023dxt}.
Mixed QCD-EW corrections have been presented in Refs~\cite{Bonetti:2020hqh,Becchetti:2021axs,Bonetti:2022lrk}, partial EW corrections are considered in Refs.~\cite{Becchetti:2018xsk,Davies:2023npk,Haisch:2024nzv}.
NNLO results for Higgs plus jet production in the HTL are available already since some time ago~\cite{Boughezal:2015dra,Boughezal:2015aha,Caola:2015wna,Chen:2016zka,Campbell:2019gmd,Chen:2019wxf,Chen:2021ibm}.
 Higgs production in association with one or multiple high-energy jets is available within the {\tt HEJ} framework~\cite{Andersen:2023kuj,Andersen:2022zte}.
In Ref.~\cite{Liu:2024tkc}, light quark mediated Higgs boson production in association with a jet at NNLO and beyond is considered in the framework of resummation.

The Higgs boson transverse momentum spectrum for boosted Higgs bosons already has been used by
the experimental collaborations to
place limits on anomalous top-Higgs and
gluon-Higgs couplings~\cite {ATLAS:2019lwq,ATLAS:2021tbi,CMS:2020zge,ATLAS:2022fnp,CMS:2024jbe}. The latter can be parameterised by $\ct$ and $\cg$ in Higgs Effective Field Theory
(HEFT), also called Electroweak Chiral Lagrangian~\cite{Feruglio:1992wf,Burgess:1999ha,Grinstein:2007iv,Contino:2010mh,Alonso:2012px,Buchalla:2013rka}.
These couplings also enter inclusive Higgs boson production in gluon fusion, which is known to agree with the SM prediction to a level approaching 5\%. Therefore, these anomalous couplings are fairly well constrained already (also from other processes such as $t\bar{t}H$ production for the case of the Higgs-top coupling~\cite{Hartland:2019bjb,Celada:2024mcf}).
However, it is well known that there is a degeneracy between $\ct$ and $\cg$ when considering only inclusive Higgs production, which is lifted when considering the $p_T$-spectrum of the Higgs boson at large transverse momenta~\cite{Grojean:2013nya,Schlaffer:2014osa,Grazzini:2016paz,Grazzini:2018eyk}. 
Up to now, the effects of these operators have not yet been studied in combination with NLO corrections to Higgs+jet production including the full top quark mass dependence.
However, both the SM top quark mass effects as well as these anomalous couplings affect the tail of the $\pth$-distribution considerably. 
Therefore it is important to study in detail the interplay of both, higher order QCD corrections and potential effects of new physics in an EFT framework.

In this work we would like to address this point and investigate the effects of these anomalous couplings on the Higgs+jet cross section and 
transverse momentum distributions, based on a calculation of the full
NLO QCD corrections in the SM~\cite{Jones:2018hbb}.
Working in HEFT rather than Standard Model Effective Field Theory (SMEFT)~\cite{Buchmuller:1985jz,Grzadkowski:2010es,Brivio:2017vri,Isidori:2023pyp}, our power counting scheme is not based on canonical dimension counting, i.e.~the counting of inverse powers of a new physics scale $\Lambda$, but instead on the counting of the chiral dimension $d_\chi$, which is related to the (explicit or implicit) loop order $L$ through $d_\chi=2L+2$.
 The chromomagnetic top-quark dipole operator has chiral dimension $d_\chi=4$ and therefore is considered to be subleading, 
 as explained in Section~\ref{sec:calculation}.
Thus the inclusion of chromomagnetic dipole operators will be considered in subsequent work, together with other subleading operators entering at the same level, such as four-fermion operators.

The structure of this paper is as follows: In Section \ref{sec:calculation}, we describe how the anomalous couplings relate to inclusive Higgs production and how they affect the large-$p_T$ spectrum of the Higgs boson. We also give some technical details about the calculation.
Section~\ref{sec:results} is dedicated to the description of phenomenological results, providing heat maps that show the effects of the anomalous couplings on the total cross section and discussing the effect of some HEFT benchmark points on the Higgs boson transverse momentum spectrum, before we conclude in Section~\ref{sec:conclusions}.

\section{Description of the method}
\label{sec:calculation}

\subsection{Framework of the calculation}


We include anomalous couplings based on the effective Lagrangian 

\begin{equation}
{\cal L} \supset -\ct\,m_t\,\frac{H}{v}\bar{t}t+\frac{\alpha_s}{8\pi}\cg\frac{H}{v}G^a_{\mu \nu}G^{a,\mu \nu}\;.
\label{eq:Lagrangian}
\end{equation}
In the SM, $\ct=1$ and $\cg=0$. We assume that the anomalous couplings
are induced by new physics interactions at a scale $\Lambda$ considerably larger than the electroweak scale.
 While the process $pp\to H+$jet is loop induced in the SM, 
 the second part of the Lagrangian now also introduces effective tree
 level interactions. The  factor proportional to 
 $\alpha_s/8\pi$ indicates that these interactions are stemming from
 loops of heavy particles that have been integrated out to arrive at
 the effective Higgs-gluon coupling.
In the region $2m_t\lesssim \sqrt{\hat{s}}\lesssim \Lambda$ the top quark loops
are resolved while the heavier particles in the loop generate the
effective point-like Higgs-gluon interaction.
The coefficient $\ct$ is a modification factor of the top Yukawa
coupling, which can arise for example by mixing with heavy top
partners~\cite{Grojean:2013nya,Banfi:2019xai}.
The chromomagnetic top-quark dipole operator ${\cal O}_{tg} = y_t g_s\bar t_L\sigma^{\mu\nu}T^aG_{\mu\nu}^a(v+H) t_R$ can only be
  generated through contracted
  loops~\cite{Arzt:1994gp,Buchalla:2022vjp,Isidori:2023pyp} in weakly coupled, renormalisable UV completions and we stick to 
such extensions of the SM. Therefore, inserting the operator into the one-loop diagrams that constitute the LO in the SM, the resulting diagrams are effectively of two-loop order (chiral dimension $d_\chi=6$), thus coming with a loop suppression factor $\sim 1/(16\pi^2)$.
The  operator ${\cal O}_{hg}$, mediating direct Higgs-gluon couplings, also pertains to the class of loop-generated operators.
However, insertions of ${\cal O}_{hg}$ enter at tree-level and therefore have chiral dimension $d_\chi=4$.
  The SM NLO QCD corrections also lead to two-loop diagrams and therefore come with a loop factor $1/(16\pi^2)$ relative to the LO diagrams.
  However, the SM two-loop QCD diagrams come with an extra factor $g_s^2=4\pi\alpha_s$ relative to the Born diagrams, while the chromomagnetic dipole operator comes with an extra factor $y_t g_s$. As we only calculate the QCD corrections, which are of order ${\cal O}(\alpha_s)$ relative to the Born amplitude, we neglect the two-loop corrections stemming from the chromomagnetic top-quark dipole operator in the present work.
  This is in line with the procedure followed in Ref.~\cite{Buchalla:2018yce} for the QCD corrections to Higgs boson pair production.
Furthermore, we do not include any 4-quark operators nor CP-violating operators.

The matrix element squared for each partonic subprocess can be written as~\cite{Schlaffer:2014osa}
\begin{equation}
\left|\mathcal{M}\right|^{2}\propto \left|\ct\,\mathcal{M}_{f}(m_{t}) + \kg\,\mathcal{M}_{Hg}\right(m_{t})|^{2},
\end{equation}
where $\mathcal{M}_{f}$ denotes the parts of the amplitude where the Higgs boson couples to a top quark, and $\mathcal{M}_{Hg}$ the amplitude parts containing an
effective point-like Higgs-gluon interaction. Note that the NLO amplitude can also contain both, a top quark loop and a 
 point-like Higgs-gluon interaction. However, in these diagrams the top quarks couple only to gluons, with an SM coupling.
 We use $\kg=\frac{3}{2}\cg$ such that the HTL corresponds to $\kg\to 1$ and $c_t=0$.
The total cross section for $pp\to H+\rm{jet}$ can be written as a
quadratic polynomial in $\ct$ and $\cg\,$, both at LO and at NLO.

Integrating over the jet momenta, the total inclusive cross section for Higgs boson production in gluon fusion should be retrieved. As is well known, due to the ``Higgs Low Energy Theorem''~\cite{Ellis:1975ap,Vainshtein:1980ea,Dawson:1989yh}, the total cross section for Higgs production in gluon fusion is rather insensitive to the masses of heavy particles circulating in the loop. This is also reflected in the fact that, 
at energy scales below $2m_t$, the inclusive Higgs production cross section
is approximated very well by the HTL.
An extra high-$p_T$ jet can serve as a handle to resolve  heavy quark loops, therefore new physics effects could show up in the tail of the Higgs $p_T$-distribution.

From the Lagrangian (\ref{eq:Lagrangian}) one obtains for  the inclusive Higgs production cross section at LO (see Ref.~\cite{Grojean:2013nya}):
\begin{align}
  \frac{\sigma_{\rm{incl}}(\ct,\kg)}{\sigma_{\rm{incl}}^{SM}}=(\ct+\kg)^2+{\cal O}(\frac{\kg}{\ct+\kg}\frac{m_H^2}{4m_t^2})\;.
  \label{eq:scale_sigtot}
\end{align}
 
As the measured total Higgs production cross section in gluon fusion
agrees very well with the SM result, the relation $(\ct+\kg)^2=1$ should
be fulfilled to about 10\% level, assuming that subleading operators do not have a drastic effect, which would lead to more freedom in the relation between $\ct$ and $\kg$. 
In the pure HTL, the proportionality of the cross section to $(\ct+\kg)^2$ is fulfilled exactly, also for the NLO amplitudes, because there are no diagrams at NLO which contain both $\ct$ and $\cg$ simultaneously, such that the HTL of the full SM NLO amplitudes gives exactly the amplitudes proportional to $\kg$. This degeneracy is broken in the Higgs boson transverse momentum spectrum because as $\pth$ increases, the top quark loops start to become resolved and therefore the kinematic behaviour of the contribution proportional to $\ct$ is different from the one in the HTL for large values of $\pth$. On the other hand, as the differential cross section decreases rapidly with $\pth$, the effects of anomalous couplings on the total cross section should be small as long as the relation $(\ct+\kg)^2=1$ is fulfilled.
Therefore it is useful to consider the cross section for $pp\to H+\rm{jet}$ as a function of the cut on $\pth$~\cite{Grojean:2013nya,Schlaffer:2014osa}:
\begin{equation}
  \label{eq:ptmin0}
  \frac{\sigma_{Hj}(\pthmin,\ct,\kg)}{\sigma_{Hj}^{\mathrm{SM}}(\pthmin)}= (\ct+\kg)^2+\delta_1(\pthmin)\, \ct\, \kg +\delta_2(\pthmin) \, \kg^2\;,
\end{equation}
where  the coefficients $\delta_i$ depend on the cut $\pthmin$.
For small $\pthmin$ the coefficients $\delta_i$ at LO are very small, modifying the cross
section in the permille to percent range below $\pth\sim 350~\giga \electronvolt$~\cite{Grojean:2013nya}.
However, recent LHC measurements have reached transverse momentum regions beyond $600~\giga \electronvolt$~\cite{CMS:2024jbe,CMS:2020zge,ATLAS:2019lwq,ATLAS:2023jdk}. Furthermore, the study of Refs.~\cite{Grojean:2013nya,Schlaffer:2014osa} was at LO only, and the one of Refs.~\cite{Grazzini:2016paz,Maltoni:2016yxb,Grazzini:2018eyk} is based on the HTL when going beyond LO.
In Section~\ref{sec:results}, we will investigate how the anomalous couplings modify the large-$\pth$ spectrum at NLO with full top quark mass dependence.

\subsection{Technical details}

The cross section for $pp\to H+$jet consists of $gg, qg, \bar{q}g$ and $q\bar{q}$ initiated subprocesses.
The calculation largely relies on the corresponding setup for the SM case, described in Ref.~\cite{Jones:2018hbb}.
 
\subsubsection*{Leading order amplitudes} 
\begin{figure}[htb]
\centering
\includegraphics[width=10cm]{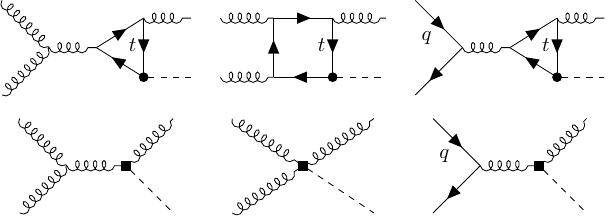}
\caption{Example diagrams contributing to Higgs plus one jet production at LO, based on the
  chiral Lagrangian given in eq.~(\ref{eq:Lagrangian}).}
\label{fig:h1j-born}
\end{figure}

The LO amplitudes in the full theory as well as the
amplitudes involving $\cg$ were implemented analytically, relying on Ref.~\cite{Baur:1989cm},
while the one-loop real radiation contribution and the two-loop virtual
amplitudes rely on semi-numerical evaluations.
As a cross-check we also generated the Born amplitudes with
\gosam~\cite{Cullen:2011ac,GoSam:2014iqq} using the UFO~\cite{Degrande:2011ua,Darme:2023jdn} model described in
Ref.~\cite{Buchalla:2018yce}, finding agreement between the two
implementations at amplitude and cross section level.
Example diagrams contributing at Born level are shown in~\Cref{fig:h1j-born}.

\subsubsection*{Real radiation}

\begin{figure}[htb]
\begin{center}
\includegraphics[width=9cm]{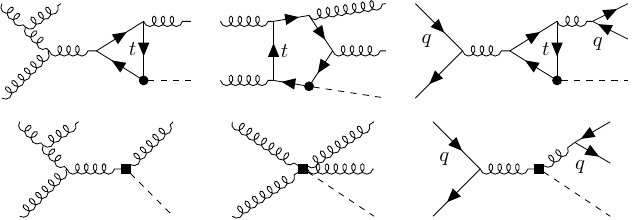}
\caption{Examples for real radiation diagrams of order $g_s^4$.}
\label{fig:h1j-real}
\end{center}
\end{figure}

The real radiation corrections contain one-loop diagrams up to pentagons
as well as tree-level 5-point-diagrams, examples are shown in~\Cref{fig:h1j-real}.
The loop-induced real radiation matrix elements were implemented
using the interface~\cite{Luisoni:2013cuh}
between \gosam~\cite{Cullen:2011ac,GoSam:2014iqq} and
the \powhegbox~\cite{Nason:2004rx,Frixione:2007vw,Alioli:2010xd}, modified
accordingly to compute the real corrections based on one-loop amplitudes for the part of the amplitude that contains explicit top quark loops. The one-loop amplitudes were generated with \gosam{}-2.0~\cite{GoSam:2014iqq}, that
uses {\tt Qgraf}~\cite{Nogueira:1991ex}, \form~\cite{Kuipers:2012rf} and
{\tt spinney}~\cite{Cullen:2010jv} for the generation of the Feynman
diagrams, and offers a choice from {\tt
Samurai}~\cite{Mastrolia:2010nb,vanDeurzen:2013pja}, {\tt
golem95C}~\cite{Binoth:2008uq,Cullen:2011kv,Guillet:2013msa}
and \ninja{}~\cite{Peraro:2014cba} for the
reduction, as well as \avholo{}~\cite{vanHameren:2010cp} or \qcdloop{}~\cite{Ellis:2007qk} for the scalar integrals.  At run time the amplitudes were computed using
\ninja{}~\cite{Peraro:2014cba}, {\tt
golem95C}~\cite{Cullen:2011kv} and \avholo{}~\cite{vanHameren:2010cp}
for the evaluation of the one-loop integrals.

\subsubsection*{Virtual corrections} 
\begin{figure}[htb]
\centering
\includegraphics[width=8cm]{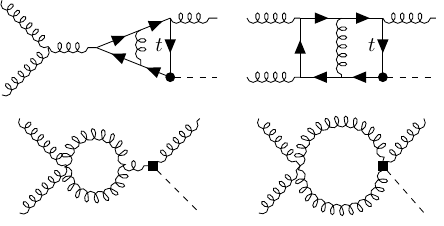}
\caption{Examples for virtual diagrams of order $g_s^5$ in the gluon fusion channel.}
\label{fig:h1j-virt}
\end{figure}

For the virtual two-loop amplitudes, we have used the results of the
calculation presented in Ref.~\cite{Jones:2018hbb},
which is based on {\sc Reduze}\,2~\cite{vonManteuffel:2012np} and {\sc SecDec}-3~\cite{Borowka:2015mxa}, which evolved to {\sc pySecDec}~\cite{Borowka:2017idc,Heinrich:2021dbf,Heinrich:2023til}.
Examples of virtual diagrams are shown in~\Cref{fig:h1j-virt}. 
The values for the Higgs boson and top quark masses have been set to
$m_H=125~\giga \electronvolt$ and $m_t=173.055~\giga \electronvolt$, which means $m_H^2/m_t^2=12/23$. Fixing these values reduces the number of independent scales in the two-loop amplitudes to two variables, the Mandelstam invariants
$\hat{s}$ and $\hat{t}$. 

The $gg \to g\,H$ amplitude can be decomposed into four tensor structures.
After imposing parity conservation, transversality of the gluon
polarization vectors and the Ward identity, the amplitude can be
written as a linear combination of four form factors $F_{ijk}$ multiplying the tensor structures $T^{\mu \nu \tau}_{ijk}$~\cite{Boggia:2017hyq}:
\begin{align}
\mathcal{M}^{\mu \nu \tau} = F_{212} T^{\mu \nu \tau}_{212}  + F_{332} T^{\mu \nu \tau}_{332} + F_{311} T^{\mu \nu \tau}_{311} + F_{312} T^{\mu \nu \tau}_{312}\;,
\end{align}
where
\begin{align}
T^{\mu \nu \tau}_{212} &= (s_{12}g^{\mu \nu}-2p_2^{\mu}p_1^{\nu})(s_{23}p_1^{\tau}-s_{13}p_2^{\tau})/(2s_{13}) \;,\nonumber \\
T^{\mu \nu \tau}_{332} &= (s_{23}g^{\nu \tau}-2p_3^{\nu}p_2^{\tau})(s_{13}p_2^{\mu}-s_{12}p_3^{\mu})/(2s_{12}) \;,\nonumber \\
T^{\mu \nu \tau}_{311} &= (s_{13}g^{\tau \mu}-2p_1^{\tau}p_3^{\mu})(s_{12}p_3^{\nu}-s_{23}p_1^{\nu})/(2s_{23}) \;,\nonumber \\
T^{\mu \nu \tau}_{312} &= \Big(g^{\mu \nu}(s_{23}p_1^{\tau}-s_{13}p_2^{\tau}) + g^{\nu \tau}(s_{13}p_2^{\mu}-s_{12}p_3^{\mu})+g^{\tau \mu}(s_{12}p_3^{\nu}-s_{23}p_1^{\nu}) \nonumber \\
&+2p_3^{\mu}p_1^{\nu}p_2^{\tau}-2p_2^{\mu}p_3^{\nu}p_1^{\tau}\Big)/2\;,
\end{align}
with $s_{ij} = (p_i+p_j)^2$. Three of the form factors are related by cyclic permutations of the external gluon momenta while the fourth is invariant under such permutations.
The $q\bar{q} \to g\,H$ amplitude similarly can be decomposed in terms of two tensor structures as~\cite{Gehrmann:2011aa}:
\begin{align}
M_{\rho} \epsilon^{\rho}= F_1 T_1+F_2 T_2\;,
\end{align}
where
\begin{align}
  T_1 &=\bar{u}(p_1)\not{p_3}\,v(p_2)p_2\cdot \epsilon_3 - \bar{u}(p_1)\not{\epsilon_3}\,v(p_2)p_2\cdot p_3\;,\nonumber \\
T_2 &=\bar{u}(p_1)\not{p_3}\,v(p_2)p_1\cdot \epsilon_3 - \bar{u}(p_1)\not{\epsilon_3}\,v(p_2)p_1\cdot p_3\;.
\end{align}
In this case the form factors are related by interchanging the external quark and anti-quark momenta.
The $qg\to q\,H$ amplitude can be obtained from the $q\bar{q} \to g\,H$ amplitude by crossing.

The form factors can be extracted introducing projectors $P^{\mu \nu \tau}_{ijk}$ satisfying $P^{\mu \nu \tau}_{ijk}M_{\mu \nu \tau}=F_{ijk}$.
The four projectors for the  $gg \to gH$ amplitude in $D$-dimensional space-time are:
\begin{align}
P_{212}^{\mu \nu \tau} &= \frac{1}{(D-3)s_{23}} \left(-\frac{Ds_{13}}{s_{12}^2s_{23}}T_{212}^{\mu \nu \tau}-\frac{D-4}{s_{23}^2}T_{332}^{\mu \nu \tau}-\frac{D-4}{s_{12}s_{13}}T_{311}^{\mu \nu \tau}+\frac{D-2}{s_{12}s_{23}}T_{312}^{\mu \nu \tau} \right)\;, \nonumber \\
P_{332}^{\mu \nu \tau} &= \frac{1}{(D-3)s_{12}} \left(-\frac{D-4}{s_{12}s_{23}}T_{212}^{\mu \nu \tau}-\frac{Ds_{12}}{s_{13}s_{23}^2}T_{332}^{\mu \nu \tau}-\frac{D-4}{s_{13}^2}T_{311}^{\mu \nu \tau}+\frac{D-2}{s_{13}s_{23}}T_{312}^{\mu \nu \tau} \right)\;, \nonumber \\
P_{311}^{\mu \nu \tau} &= \frac{1}{(D-3)s_{13}} \left(-\frac{D-4}{s_{12}^2}T_{212}^{\mu \nu \tau}-\frac{D-4}{s_{13}s_{23}}T_{332}^{\mu \nu \tau}-\frac{Ds_{23}}{s_{12}s_{13}^2}T_{311}^{\mu \nu \tau}+\frac{D-2}{s_{12}s_{13}}T_{312}^{\mu \nu \tau} \right)\;, \nonumber \\
P_{312}^{\mu \nu \tau} &= \frac{D-2}{(D-3)s_{12}s_{13}s_{23}} \left(\frac{s_{13}}{s_{12}}T_{212}^{\mu \nu \tau}+\frac{s_{12}}{s_{23}}T_{332}^{\mu \nu \tau}+\frac{s_{23}}{s_{13}}T_{311}^{\mu \nu \tau}+\frac{D}{D-2}T_{312}^{\mu \nu \tau} \right)\;.  \label{eq:projgg}
\end{align}
For the $q\bar{q}\to gH$ amplitude the projectors are~\cite{Gehrmann:2011aa}:
\begin{align}
P_1&=\frac{D-2}{2(D-3)s_{12}s_{13}^2}T_1^{\dag}-\frac{D-4}{2(D-3)s_{12}s_{13}s_{23}}T_2^{\dag} \;,\nonumber \\
P_2&=-\frac{D-4}{2(D-3)s_{12}s_{13}s_{23}}T_1^{\dag}+\frac{D-2}{2(D-3)s_{12}s_{23}^2}T_2^{\dag}\;,\label{eq:projqq}
\end{align}
where $P_1$ and $P_2$ satisfy  $\sum_{\rm{spins}} P_i M_{\rho}\epsilon^{\rho}=F_i.$
The six NLO QCD form factors have been computed for the SM case in Ref.~\cite{Jones:2018hbb}.
We have rescaled the SM form factors by the anomalous coupling $c_t$.
To construct the virtual corrections, we use 
\begin{align}           
d\sigma^\mathrm{V} &\sim 2\Re\left(\mathcal{M}^\mathrm{V}\cdot\mathcal{M}^{\mathrm{B},\dag}\right)=2\Re\left[\left(\mathcal{M}^{\mathrm{2L}}_{f}+\mathcal{M}^{\mathrm{1L}}_{Hg}\right)\cdot\left(\mathcal{M}^{\mathrm{1L}}_{f}+\mathcal{M}^{\mathrm{0L}}_{Hg}\right)^{\dag}\right] \nonumber \\
&=2\Re\left[\mathcal{M}^{\mathrm{2L}}_{f}\cdot\left(\mathcal{M}^{\mathrm{1L}}_{f} + \mathcal{M}^{\mathrm{0L}}_{Hg}\right)^\dag+\mathcal{M}^{\mathrm{1L}}_{Hg}\cdot\left(\mathcal{M}^{\mathrm{1L}}_{\mathrm{f}}+\mathcal{M}^{\mathrm{0L}}_{Hg}\right)^\dag\right]\;.
\end{align}
The Born matrix elements arising from the tree level diagrams where the Higgs boson couples to gluons, $\mathcal{M}^{\mathrm{0L}}_{Hg}$, are added  to the rescaled SM Born matrix elements at form factor level, using the projectors of eqs.~(\ref{eq:projgg}) and (\ref{eq:projqq}). 

In order to use the 2-loop virtual contribution directly within the \powhegbox{}, we constructed a grid based on the 2000 phase space points  at which the 2-loop amplitude has been evaluated, together with an interpolation framework.
To this aim we first realise that we can write the amplitude as
\begin{equation}
     M(\mu) = M_2\log(\mu)^2 + M_1\log(\mu)+M_0.
 \end{equation}
 Thus, we computed the amplitudes for three values of $\mu$ and then solved the equations for $M_0,M_1$ and $M_2$. 
 We created a grid for each of the four partonic channels where we stored
	\begin{align}
		\beta=\frac{s-m_H^2}{s+m_H^2},~\cos \theta=\frac{t-u}{s-m_H^2}, ~M_0, M_1, M_2.
	\end{align}
 To interpolate the virtual amplitudes for each channel we used
	a neural network trained on the grid, where we used 30\% of the data points as a validation set. For the training of the network we multiplied the coefficients $M_0,M_1$ and $M_2$ with $\beta^2(1-\beta)(1-\cos^2\theta)$ and divided by the largest value in the grid. This guarantees that the grid for the neural network training has values in $[-1,1]$ and is flattened at the phase space boundaries.  
	The architecture of each network is composed of three dense layers with
	200, 20 and 20 nodes, respectively. The setup is implemented using {\tt Keras}~\cite{chollet2015keras}  to produce the models in combination with a modified version of {\tt Keras2cpp}~\cite{keras2cpp} to save them such that they can be loaded from C++. We
	built a C++ function where, for each channel, the models are loaded. This function is called in the
	\powhegbox{} to obtain the two-loop virtual amplitude at any given
	phase space point. 
 
\par
The one-loop amplitudes contributing to the virtual corrections where the Higgs couples to gluons, denoted by  $\mathcal{M}^{\mathrm{1L}}_{Hg}$, were computed by \gosam{}, where the interference $\mathcal{M}^{\mathrm{1L}}_{Hg}\cdot\mathcal{M}^{\mathrm{0L}}_{Hg}$ is straightforward, while we had to slightly modify \gosam{} in order to compute the $\mathcal{M}^{\mathrm{1L}}_{Hg}\cdot\mathcal{M}^{\mathrm{1L}}_{f}$ interference.

One full run takes roughly 320 CPU-days, depending on the used hardware. The 320 CPU-days refer to a cluster
which consists of Intel Xeon Gold 6230 processors with a frequency of 2.1 GHz. It is also worth mentioning that using either $\ct=0$ or $\cg=0$ can drastically reduce the number of CPU-hours, since this greatly reduces the number of diagrams to be evaluated.

\subsubsection*{Validation} 
In order to allow for comparisons and cross checks, we implemented
both the $m_t\to\infty$ limit as well as the full SM amplitudes at
NLO. We checked that taking $m_t\to\infty$ in all diagrams 
and setting $c_g=0$ agrees with the SM calculation in the HTL. Furthermore, using the fact that in the HTL the SM reduces to diagrams with an effective Higgs-gluon coupling given by $c_{g,\mathrm{HTL}}=2/3$, see e.g. \cite{Spira:1995rr}, and that the HEFT diagrams with a gluon-Higgs coupling reduce to just HTL diagrams without any top-loops, we checked that $\ct\mathcal{M}_{f}(m_t\to\infty)+\cg\mathcal{M}_{Hg}(m_t\to\infty)=(2/3\ct+\cg)/c_{g,\mathrm{HTL}}\cdot\mathcal{M}_{\mathrm{HTL}}$.  
Furthermore, we have validated the interpolation grid that we implemented in the {\tt Powheg} setup by comparing the grid-based result with the result reconstructed directly from the points obtained from the numerical evaluation of the two-loop amplitude. The differences were of the order of the Monte Carlo uncertainties. We also interpolated the Born amplitude of the grid and compared it to the analytical results.
Of course, we also checked that taking $c_t=1$ and $c_g=0$ agrees with the SM results computed in \cite{Jones:2018hbb,Chen:2021azt}, both at amplitude level and at total cross section level.

\section{Numerical results and discussion of anomalous couplings}
\label{sec:results}

The results presented in this section were obtained using the
PDF4LHC21{\tt\_}40{\tt\_pdfas} parton distribution functions~\cite{Butterworth:2015oua,Dulat:2015mca,Harland-Lang:2014zoa,NNPDF:2014otw}
 interfaced to our code via
LHAPDF~\cite{Buckley:2014ana}, along with the corresponding value for
$\alpha_s$.  The masses of the Higgs boson and the top quark have been
fixed, as in the virtual amplitude, to $m_H=125~\giga \electronvolt$, $m_t=173.055~\giga \electronvolt$,
respectively. Their widths have been set to zero.   
Jets are clustered with the
anti-$k_T$ algorithm~\cite{Cacciari:2008gp} as implemented in the
{\tt fastjet} package~\cite{Cacciari:2005hq, Cacciari:2011ma}, with jet
radius $R=0.4$ and a minimum transverse momentum 
$p_{T,\rm{min}}^{\rm{jet}}=30~\giga \electronvolt$.
The central scale is given by
\begin{equation}
    \mu_0 =H_T/2=\frac{1}{2}\left(\sqrt{m_H^2+\pth^2}+\sum_i |p_{T,i}|\right), 
\end{equation}
 where the sum is over all final state partons $i$. The scale uncertainties are
estimated by varying the factorisation and renormalisation scales $\muF$ and  $\muR$. The uncertainty bands represent the envelopes of a 3-point scale variation around the central scale.

\subsection{Total cross sections and heat maps}

\begin{figure}[htb]
\includegraphics[]{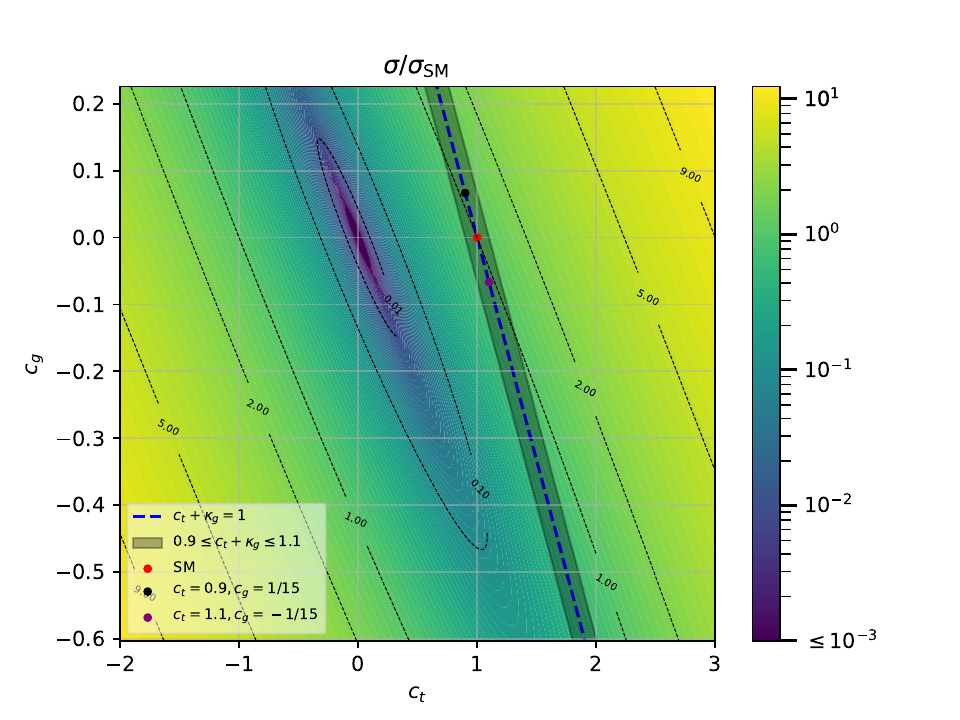}
\caption{Heat map of the ratio of the HEFT to the SM at the central scale. The colour spectrum is capped at $10^{-3}$. The cut on $\pth$ is $400~\giga\electronvolt$.} \label{fig:heatmap1}
\end{figure}

\begin{figure}[htb]
\includegraphics[]{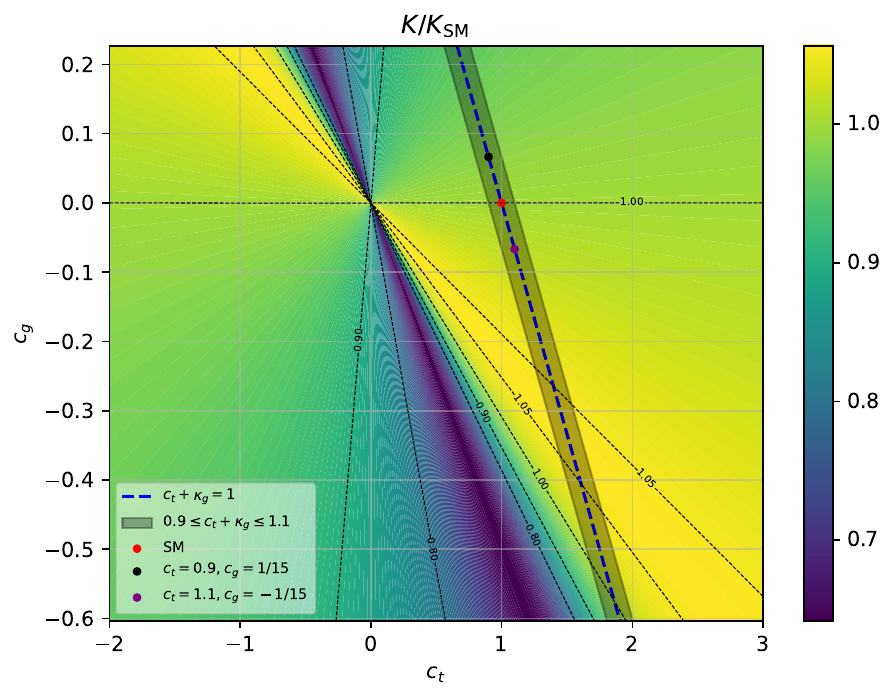}
\caption{Heat map of the ratio of the NLO K-factors of the HEFT to the SM at the central scale.  The cut on $\pth$ is $400~\giga\electronvolt$.\label{fig:heatmap2}}
\end{figure}

The total cross sections obtained with the above settings and for different values of the minimum transverse momentum of the Higgs boson, $\pth^{\rm{min}}$, are given in Table~\ref{tab:totalxs}, where we compare the benchmark point $\ct=0.9, \cg=1/15$ to the SM case. We can clearly see that at LO, the difference between BSM and SM is within the corresponding scale uncertainties. At NLO, the scale uncertainties are reduced and the difference becomes noticeable for highly boosted Higgs bosons, but only for a $\pth$ cut of $800\,\giga \electronvolt$ the difference is clearly outside the scale uncertainties for the considered benchmark point. This shows that already for small deviations from the SM there can be a measurable difference, if the Higgs boson is very highly boosted. Going away further from the SM values for $\cg$ and $\ct$, the difference would become more pronounced and start at a smaller $\pth$--cut values. This will become apparent in the discussion of the heat maps shown in Figs.~\ref{fig:heatmap1} and \ref{fig:heatmap2}.
In Table~\ref{tab:Ratio_to_SM}, we show the ratio of the HEFT at the benchmark point $\ct=0.9$, $\cg=1/15$ to the SM case, including Monte Carlo uncertainties rather than scale uncertainties, as the scale uncertainties are correlated between HEFT and the SM and therefore largely cancel.
The ratios clearly demonstrate the behaviour discussed above, i.e.\ the difference between HEFT and the SM becoming more pronounced for higher values of $\pth^{\rm{min}}$.

\begin{table}
\centering
    \begin{tabular}{ |c||c|c|c|c| }
        \hline
        \multirow{2}{*}{$\pth$ cut [$\giga \electronvolt]$} & \multicolumn{2}{c|}{$\sigma_\mathrm{cut}$(HEFT) [fb]} & \multicolumn{2}{c|}{$\sigma_\mathrm{cut}$(SM) [fb]}  \\
        & LO & NLO & LO & NLO \\
        \hline
        \hline
        0 &$(8\substack{+3 \\ -2})\cdot10^{3}$ & $ (15\substack{+0 \\ -2})\cdot10^{3} $ & $(8\substack{+3\\ -2})\cdot10^{3} $ &$(15\substack{+0\\ -2})\cdot10^{3} $  \\
        \hline
        50  & $(4.5\substack{+1.8  \\ -1.2})\cdot10^{3} $ & $(9.1\substack{+0.1 \\ -1.6})\cdot10^{3}  $ & $(4.5\substack{+1.8 \\ -1.2})\cdot10^{3}$ & $(9.1\substack{+0.0 \\ -1.4})\cdot10^{3} $\\
        \hline
        100 &$(1.4\substack{+0.6 \\ -0.4})\cdot10^{3}$  & $(2.85\substack{+0.08 \\ -0.5})\cdot10^{3}$ & $(1.4\substack{+0.6 \\ -0.4})\cdot10^{3} $ & $(2.85\substack{+0.01 \\ -0.5})\cdot10^{3} $    \\
        \hline
        200 & $(2.2\substack{+0.9 \\ -0.6})\cdot10^{2}$ & $(4.6\substack{+0.2 \\ -0.9})\cdot10^{2}$  &$(2.2 \substack{+0.9 \\ -0.6})\cdot10^{2}$ & $(4.5\substack{+0.1 \\ -0.9}) \cdot10^{2}$  \\
        \hline
        400 &$ 13\substack{+6 \\ -4}$ & $28\substack{+2 \\ -6}$ & $12\substack{+5 \\ -3}$ & $25\substack{+1.6 \\ -5}$   \\
        \hline
        600 & $1.6\substack{0.7 \\ -0.5}$  & $3.3 \substack{+0.3\\ -0.7}$   & $1.3\substack{+0.6 \\ -0.4}$  & $2.7\substack{+0.2 \\ -0.5}$  \\
        \hline
        800 & $0.29\substack{+0.14 \\-0.09 }$  & $0.60\substack{+0.07\\-0.12 }$   & $0.21\substack{+0.10 \\ -0.06}$  & $0.43\substack{+0.04 \\-0.09 }$  \\
        \hline
    \end{tabular}
\caption{Total cross sections for different values of $\pth^{\rm{min}}$, The main value is based on the central scale and the uncertainties are obtained from three-point variations around the central scale. The HEFT values are for the benchmark point $\ct=0.9, \cg=1/15$. All cross sections have a minimal jet-$p_\mathrm{T}$ of $30~\giga\electronvolt$ and thus at LO, the Higgs boson also has a minimal transverse momentum of $30~\giga\electronvolt$.} 
\label{tab:totalxs}
\end{table}

\begin{table}
\centering
    \begin{tabular}{ |c||c|c|c| }
        \hline
        \multirow{2}{*}{$\pth$ cut [$\giga \electronvolt]$} & \multicolumn{2}{c|}{${\sigma_\mathrm{cut.HEFT}}/{\sigma_\mathrm{cut,SM}}$} \\
        & LO & NLO\\
        \hline
        \hline
        0 &$0.9955 \pm 0.0005$ &$1.00 \pm 0.04$ \\\hline
        50 & $0.9966 \pm 0.0006$ & $1.00 \pm 0.04$ \\\hline
        100 & $1.0002 \pm 0.0010$ &$1.00 \pm 0.02$\\\hline
        200 & $1.021 \pm 0.002$ &$1.01 \pm 0.014$\\\hline
        400 & $1.119 \pm 0.007$ &$1.10\pm 0.01$\\\hline
        600 & $1.250 \pm 0.012$&$1.22 \pm 0.01$\\\hline
        800 & $1.410 \pm 0.016$ &$1.37 \pm 0.01$\\\hline
    \end{tabular}
\caption{The ratios of the total cross section for different values of $\pth^{\mathrm{min}}$ for a HEFT benchmark point with $(\ct,\cg)=(0.9,1/15)$ relative to the SM cross section. The uncertainties are the error propagated Monte Carlo uncertainties, since the scale uncertainties are correlated in HEFT and SM and thus mostly cancel. }
\label{tab:Ratio_to_SM}
\end{table}
The heat maps illustrate the effects of varying $c_t$ and $c_g$ simultaneously, over a parameter range inspired by current bounds from global fits~\cite{Ethier:2021bye,Celada:2024mcf}.
In Fig.~\ref{fig:heatmap1}, we show the ratio of the total cross section including anomalous couplings within HEFT to the SM total cross section, both calculated at full NLO. 
In Fig.~\ref{fig:heatmap2}, we show how the NLO K-factor is modified by the anomalous couplings.

For the heat maps, we write the total cross section as
\begin{equation}\label{eq:ABC}
    \sigma = \ct^2 A + 2\ct\cg \, B+\cg^2 \, C \; , \quad\mbox{with}\;\sigma_{\mathrm{SM}} = A\; , \quad \sigma_{\mathrm{HTL}} = \left(\frac{2}{3}\right)^2 C_{\mathrm{HTL}}\;,
\end{equation}
where the last relation is based on the fact that, if we denote the effective gluon-Higgs coupling in the HTL of the SM as $c_{g,\mathrm{HTL}}$, then the cross section can be written as $\sigma_{\rm{HTL}}=c_{g,\rm{HTL}}^2C_{\rm{HTL}}$, with $c_{g,\rm{HTL}}=2/3$ as explained above.
Thus it is sufficient to compute the cross section for different values of $\cg$ and $\ct$ and then fit the coefficients $A,B$ and $C$. For the fit we chose some ($\ct,\cg$) value pairs across a wide range, even outside the experimental limits, in order to guarantee a good fit.

By construction $A,B,C$ are independent of the variables $\ct,\cg$ and thus
\begin{align}
   \frac{\sigma}{\sigma_{\mathrm{SM}}} &= \ct^2 + 2\ct\cg \,\frac{B}{A} + \cg^2 \,\frac{C}{A},\label{eq:ratio_to_SM}\\
    \frac{\sigma_{\mathrm{NLO}}}{\sigma_{\mathrm{LO}}} &=  \frac{\ct^2 A_{\mathrm{NLO}}+2\ct\cg B_\mathrm{NLO}+\cg^2C_\mathrm{NLO}}{\ct^2 A_{\mathrm{LO}}+2\ct\cg B_\mathrm{LO}+\cg^2C_\mathrm{LO}}\;.\label{eq:ratio_to_LO}
\end{align}
Hence it is sufficient to first perform a fit of the LO and NLO coefficients and then use equations \eqref{eq:ratio_to_SM} and \eqref{eq:ratio_to_LO} to compute the K-factors and the ratio to the SM.

\par

In Table~\ref{tab:ABC_fit}, we list the fitted values for the coefficients using different cuts on the Higgs boson transverse momentum. The corresponding values for variations around the central scale are given in appendix~\ref{sec:appendix}.

We see that for all cuts the largest coefficient is always the $C$-coefficient, followed by  $B$ and then $A$. Thus the dominant contributions to the cross sections stem from purely HTL-like diagrams. 
Note that the reason why for the chosen benchmark point the SM-like diagrams are more important comes from the fact that $\ct=0.9\gg\cg=1/15$, compatible with current constraints.  Furthermore, the higher the cut on the transverse momentum of the Higgs boson, the bigger the ratios $C/A$ and $B/A$ become, indicating again that the deviations from the SM start to become more important for high $\pth$. This stems from the fact that $d\sigma/d\pth \sim 1/\pth^a$ scales with $a = 2$ in the
full theory and with $a = 1$ in the HTL~\cite{Caola:2016upw}.
We can also see that $C_{\mathrm{HTL}} \approx C$ for all values of the cut. This is to be expected since the HTL in the SM corresponds to the HEFT with $\cg=2/3$, excluding those top-loops where the top quark only couples to gluons.  The latter are suppressed in the HTL since they do not involve a Yukawa coupling. Thus this gives a cross-check of our computations.
\par

\begin{table}[bt]
    \centering
    \begin{tabular}{ |c|c|c|c|c|c| }
        \hline
        {$p_{T,H}^\mathrm{cut}$} &{$A$} & {$B$} & {$C$} & {$C_\mathrm{HTL}$}\\\hline
         0 & $15.0 \pm 0.2$ &$21.4 \pm 0.5$ &  $ 30.7\pm0.8 $&$30.7 \pm 0.8$\\\hline
         50 & $9.13\pm 0.12$& $13.1\pm 0.3 $ & $18.9\pm 0.5$ & $18.3\pm 0.5$ \\\hline
         100 & $2.86 \pm 0.02$&  $4.16 \pm 0.07$ &  $ 6.17 \pm 0.1$ & $6.13 \pm 0.09$ \\\hline
         200 & $(4.54 \pm 0.03)\cdot 10^{-1}$&$(7.16 \pm 0.12)\cdot 10^{-1}$& $ (11.9\pm 0.2)\cdot 10^{-1}$ & $(11.9 \pm 0.1)\cdot 10^{-1}$ \\\hline
         400 & $(2.54 \pm 0.01)\cdot 10^{-2}$  & $(5.65 \pm 0.08)\cdot 10^{-2}$ &$(13.7\pm 0.1)\cdot 10^{-2}$ & $(13.6 \pm 0.1) \cdot 10^{-2}$\\\hline
         600 & $(2.73 \pm 0.01) \cdot 10^{-3}$& $(8.37 \pm  0.11)\cdot 10^{-3}$& $(28.9\pm 0.2)\cdot 10^{-3}$ & $(28.6 \pm 0.1)\cdot 10^{-3}$\\\hline
         800 & $(4.33 \pm 0.02)\cdot 10^{-4}$ & $(17.5 \pm 0.2)\cdot 10^{-4}$ & $(80.6 \pm 0.5) \cdot 10^{-4}$ & $(80.0 \pm 0.3)\cdot 10^{-4}$ \\\hline
    \end{tabular}
    \caption{Values for $A$,$B$ and $C$ at NLO in $\pico\barn$ at the central scale. The values for the $\pth$--cut are given in $\giga \electronvolt$. The uncertainties are the uncertainties of the fit.}\label{tab:ABC_fit}
\end{table}

As can be seen in \Cref{fig:heatmap1}, the difference to the SM cross section can get very pronounced as we deviate further from the interval $\ct+3/2\cg \in [0.9,1.1]$. We can also see that the values of $(\ct,\cg)=(0.9,1/15)$ and $(\ct,\cg)=(1.1,-1/15)$ are at the boundary to where a difference to the SM starts to become significant. We find that $66.67\%$ of the HEFT points with $\ct+3/2\cg \in [0.9,1.1]$ deviate from the SM result by up to $38\%$. If we enlarge the interval to $[0.8,1.2]$, the deviation for 2/3 of the HEFT points in that interval increases to $41\%$.

In \Cref{fig:heatmap2} we show the ratio of the HEFT K-factors to the SM K-factor. We see that the relative K-factors vary significantly as a function of $\ct$ and $\cg$.


\subsection{Higgs boson transverse momentum distributions}

\begin{figure}[htb]
\centering
\includegraphics[width=\textwidth]{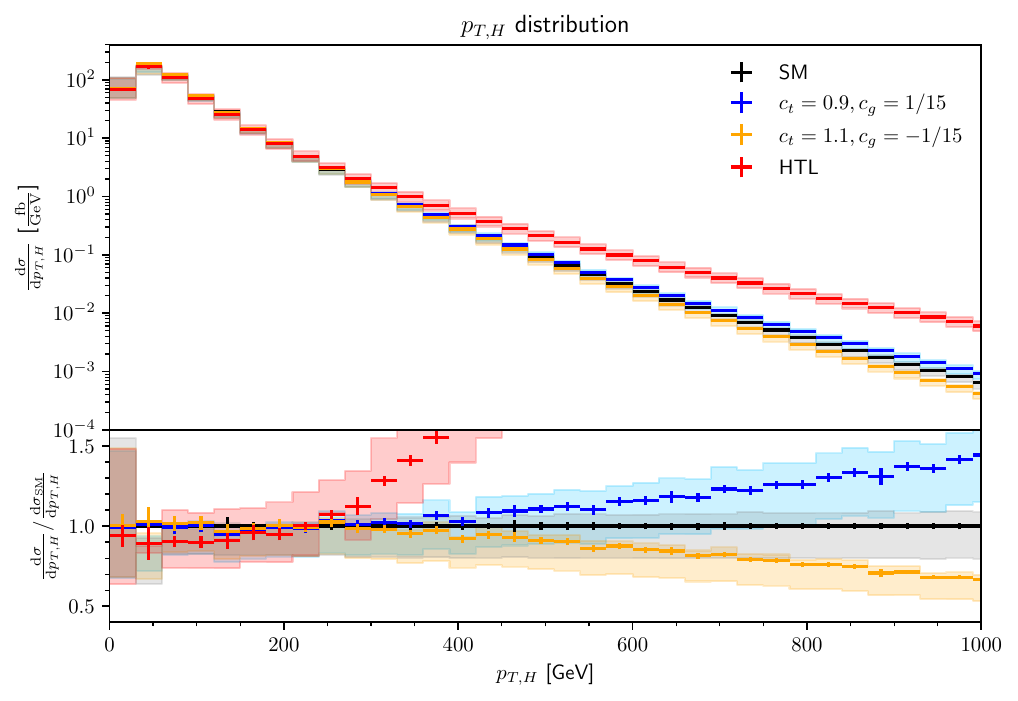}
\caption{Higgs boson transverse momentum distribution for two HEFT benchmark points, $(\ct,\cg)=(0.9,1/15)$ and $(\ct,\cg)=(1.1,-1/15)$, compared to the SM case and to the HTL. The bands denote 3-point scale variations around the central scale $\mu_0 =H_T/2$.\label{fig:Hpt}}
\end{figure}

The $\pth$ distribution with a minimum $p_T$-cut of $30\,\giga\electronvolt$ on the jet is shown in \Cref{fig:Hpt}, for the SM, the HTL and two HEFT benchmark points. Up to values of $\pth\approx300~\giga\electronvolt$ there is no significant difference between the considered predictions, but for larger $\pth$ values the HTL shows large deviations from the SM, whereas the two HEFT parameter points lead to significantly smaller deviations. Thus it is very important to use the SM predictions with full $m_t$-dependence, otherwise the approximation given by the HTL could mask enhancements in the tail which are in fact BSM effects.
\par
For the two benchmark points we consider, the BSM effects only lie outside the SM NLO QCD scale uncertainty bands for $\pth\gtrsim 800~\giga\electronvolt$. However, the two benchmark points we chose are quite close to the SM. Choosing larger deformations of the SM case would lead to more pronounced effects, visible already at smaller $\pth$ values. 
Nonetheless, for very highly boosted Higgs bosons, even small deviations from the SM couplings can lead to characteristic effects. 
It remains to be investigated whether other SM uncertainties, such as the choice of different top mass renormalisation schemes, can lead to shape distortions that could mask BSM effects. 
In Ref.~\cite{Bonciani:2022jmb} it was shown that the $\pth$ distribution with the top quark mass renormalised in the $\overline{\mathrm{MS}}$ scheme falls off faster than in the on-shell (OS) scheme as $\pth$ increases. However, the ratio OS/$\overline{\mathrm{MS}}$ in the $\pth$ spectrum stays rather constant for $\pth$ values between 600\,\giga\electronvolt and 1\,T\electronvolt, while the BSM effects grow much more rapidly with $\pth$.\\
Similar considerations hold for the QCD corrections beyond NLO. In Ref.~\cite{Becker:2020rjp} the NLO K-factors have been shown to be rather uniform over the whole $\pth$ spectrum, both in the full SM as well as for the HTL. For the case of the HTL, the ratio between NNLO and NLO also turned out to be rather flat, NNLO increasing the NLO corrections by about 25\% for $400\,\giga\electronvolt \leq \pth\leq
1\,\tera\electronvolt$. Thus, a distinctive feature of the anomalous couplings consists in the rapid growth of the shape distortion compared to the SM as $\pth$ increases.

\section{Conclusions}

We have presented results for Higgs boson production in association with one jet, combining full NLO QCD corrections with  the leading operators within Higgs Effective Field Theory (HEFT), which can modify the top-Higgs Yukawa coupling and induce an effective Higgs-gluon coupling.
We have taken into account constraints arising from the fact that inclusive Higgs production measurements show very good agreement with the SM prediction, assuming that subleading operators, such as the chromomagnetic operator or four-fermion operators, do not play a substantial role.
We found that there are combinations of $\ct$ and $\cg$ that reproduce the inclusive Higgs production cross section at the 10\% level, while changing the Higgs + jet cross section to an extent that exceeds the scale uncertainties in the highly boosted Higgs regime,  i.e. for $\pthmin \gtrsim 600~\giga \electronvolt$.
We also found that the NLO K-factor can vary by more than 30\% compared to the SM K-factor within the allowed range of $\ct$ and $\cg$. Furthermore, we showed that including the full top quark mass dependence is important to avoid that the approximation given by the HTL, leading to an enhancement in the tail of the $\pth$ distribution, is masking a BSM effect.

A more realistic study should also take into account decays of the Higgs boson as well as prospective statistical uncertainties.
Furthermore, it remains to be investigated in more detail whether top mass renormalisation scheme uncertainties could swamp the effects of anomalous couplings even for highly boosted Higgs bosons. On the other hand, a better control of the top mass scheme uncertainties along the lines of Ref.~\cite{Jaskiewicz:2024xkd} seems feasible.
In addition, the results presented in Ref.~\cite{Bonciani:2022jmb} for the SM case suggest that the ratio between on-shell and $\overline{\mathrm{MS}}$ renormalisation schemes for the top quark is rather flat for large $\pth$ values, while the BSM effects grow rapidly with $\pth$.

Similar considerations hold for QCD corrections beyond NLO, as the K-factors are close to constant in the large $\pth$ range. This holds for the NLO K-factors both in the HTL and in the full SM, as well as for K$_{\rm{NNLO}}$ in the HTL~\cite{Becker:2020rjp}.
The effects of subleading operators such as the chromomagnetic dipole operator or four-fermion operators also deserve further study, as well as electroweak corrections. \\
 The code is available from the \powhegbox\ website~\cite{powheg-box} under the process name {\tt Hjet\_full\_mt}.

\label{sec:conclusions}

\section*{Acknowledgements}
We would like to thank Matteo Capozi for collaboration at earlier stages of this project.
We are also grateful to Gerhard Buchalla, Stephen Jones  and Ludovic Scyboz for helpful discussions.
This research was supported by the Deutsche Forschungsgemeinschaft (DFG, German Research Foundation) under grant 396021762 - TRR 257. Parts of the computation were carried out on the BwUniCluster2, thus the authors acknowledge support by the state of Baden-Württemberg through bwHPC.

\renewcommand \thesection{\Alph{section}}
\setcounter{section}{0}
\setcounter{equation}{0}

\section{Coefficients of coupling structures}
\label{sec:appendix}

In this appendix we list the coefficients $A,B$ and $C$ of the anomalous coupling structures, see \Cref{eq:ABC},  for the three scale choices $\muR=\muF=\mu_0$, $\muR=\muF=2\mu_0$ and $\muR=\muF=\mu_0/2$. The uncertainties are the uncertainties from the fit.

\begin{table}[hbt]
    \centering
    \begin{tabular}{ |c|c|c|c|c| }
        \hline
        {$\pth$ cut [$\giga \electronvolt]$} & {$\muR=\muF=\mu_0$} & {$\muR=\muF=2\mu_0$} & {$\muR=\muF=0.5\mu_0$} \\\hline
         0 & $15.0 \pm 0.2$ & $12.6 \pm 0.13 $ &  $14.2 \pm 0.3$\\\hline
         50 & $9.13\pm 0.12$& $7.53 \pm 0.08$ & $8.98 \pm 0.21$ \\\hline
         100 & $2.86 \pm 0.02$& $2.35 \pm 0.02$ & $2.88 \pm 0.04$\\\hline
         200 & $(4.54 \pm 0.03)\cdot 10^{-1}$& $(3.69\pm 0.02)\cdot 10^{-1}$& $(4.67\pm 0.04)\cdot 10^{-1}$ \\\hline
         400 & $(2.54 \pm 0.01)\cdot 10^{-2}$  & $(2.04 \pm 0.01)\cdot 10^{-2}$ &$(2.70 \pm 0.02)\cdot 10^{-2}$ \\\hline
         600 & $(2.73 \pm 0.01) \cdot 10^{-3}$& $(2.19 \pm 0.01)\cdot 10^{-3}$& $(2.94\pm 0.02)\cdot 10^{-3}$ \\\hline
         800 & $(4.33 \pm 0.02) \cdot 10^{-4}$ & $(3.46 \pm 0.01) \cdot 10^{-4}$ & $(4.70 \pm 0.03)\cdot 10^{-4}$ \\\hline
    \end{tabular}
    \caption{Values for A at NLO in $\pico\barn$}\label{tab:A_fit}
\end{table}
\vspace{-0.25cm}
\begin{table}[hbt]
    \centering
    \begin{tabular}{ |c|c|c|c|c| }
        \hline
        {$\pth$ cut [$\giga \electronvolt]$} & {$\muR=\muF=\mu_0$} & {$\muR=\muF=2\mu_0$} & {$\muR=\muF=0.5\mu_0$} \\\hline
         0 & $21.4 \pm 0.5$ & $18.0 \pm 0.3$ & $22.6 \pm 0.8$\\\hline
         50 & $13.1\pm 0.3 $& $ 10.7\pm 0.2 $ & $ 14.4\pm 0.5$ \\\hline
         100 & $4.16 \pm 0.07$& $3.40 \pm 0.05 $ & $4.62 \pm 0.12$\\\hline
         200 & $(7.16 \pm 0.12)\cdot 10^{-1}$& $(5.81\pm 0.08)\cdot 10^{-1}$& $(8.05\pm 0.19)\cdot 10^{-1}$ \\\hline
         400 & $(5.65 \pm 0.08)\cdot 10^{-2}$  & $(4.54 \pm 0.05)\cdot 10^{-2}$ &$(6.46 \pm 0.13 )\cdot 10^{-2}$ \\\hline
         600 & $(8.37 \pm  0.11) \cdot 10^{-3}$& $(6.70 \pm 0.07 )\cdot 10^{-3}$& $(9.62 \pm 0.18 )\cdot 10^{-3}$ \\\hline
         800 & $(17.5 \pm 0.2) \cdot 10^{-4}$ & $(14.0 \pm 1.0) \cdot 10^{-4}$ & $(20.1 \pm 0.3)\cdot 10^{-4}$ \\\hline
    \end{tabular}
    \caption{Values for B at NLO in $\pico\barn$}\label{tab:B_fit}
\end{table}
\vspace{-0.25cm}
\begin{table}[hbt]
    \centering
    \begin{tabular}{ |c|c|c|c|c| }
        \hline
        {$\pth$ cut [$\giga \electronvolt]$} & {$\muR=\muF=\mu_0$} & {$\muR=\muF=2\mu_0$} & {$\muR=\muF=0.5\mu_0$} \\\hline
         0 & $ 30.7\pm 0.8 $ & $25.8 \pm 0.5$ &  $36.3\pm 1.4 $\\\hline
         50 & $18.9\pm 0.5$& $15.4 \pm 0.3$ & $23.0 \pm 0.8$ \\\hline
         100 & $ 6.17 \pm 0.1$& $5.03 \pm 0.08$ & $7.52 \pm 0.2 $\\\hline
         200 & $ (11.9\pm 0.2)\cdot 10^{-1}$& $(9.74\pm  0.01)\cdot 10^{-1}$& $(14.4\pm 0.3)\cdot 10^{-1} $ \\\hline
         400 & $( 13.7\pm 0.1)\cdot 10^{-2}$  & $( 11.2\pm 0.1)\cdot 10^{-2}$ &$(16.6 \pm 0.2)\cdot 10^{-2}$ \\\hline
         600 & $( 28.9\pm 0.2 ) \cdot 10^{-3}$& $(23.4 \pm 0.2)\cdot 10^{-3}$& $(34.9\pm 0.4)\cdot 10^{-3}$ \\\hline
         800 & $(80.6 \pm 0.5) \cdot 10^{-4} $ & $(65.2 \pm 0.3) \cdot 10^{-4}$ & $(97.5 \pm 1.0) \cdot 10^{-4}$ \\\hline
    \end{tabular}
    \caption{Values for C at NLO in $\pico\barn$}\label{tab:C_fit}
\end{table}

\bibliographystyle{JHEP}
 
\bibliography{refs_Hj}

\end{document}